\documentclass{pasj00}

\newcounter{author}
\def\authorcount#1#2{\refstepcounter{author}\label{#1}
                     \altaffiltext{\ref{#1}}{#2}}

\begin{document}
\SetRunningHead{T. Kato et al.}{ASASSN-14cc: Helium Analog of RZ LMi}

\Received{201X/XX/XX}
\Accepted{201X/XX/XX}

\title{ASASSN-14cc: Likely Helium Analog of RZ Leonis Minoris}

\author{Taichi~\textsc{Kato},\altaffilmark{\ref{affil:Kyoto}*}
        Franz-Josef~\textsc{Hambsch},\altaffilmark{\ref{affil:GEOS}}$^,$\altaffilmark{\ref{affil:BAV}}$^,$\altaffilmark{\ref{affil:Hambsch}}
        Berto~\textsc{Monard},\altaffilmark{\ref{affil:Monard2}}
}

\authorcount{affil:Kyoto}{
     Department of Astronomy, Kyoto University, Kyoto 606-8502}
\email{$^*$tkato@kusastro.kyoto-u.ac.jp}

\authorcount{affil:GEOS}{
     Groupe Europ\'een d'Observations Stellaires (GEOS),
     23 Parc de Levesville, 28300 Bailleau l'Ev\^eque, France}

\authorcount{affil:BAV}{
     Bundesdeutsche Arbeitsgemeinschaft f\"ur Ver\"anderliche Sterne
     (BAV), Munsterdamm 90, 12169 Berlin, Germany}

\authorcount{affil:Hambsch}{
     Vereniging Voor Sterrenkunde (VVS), Oude Bleken 12, 2400 Mol, Belgium}

\authorcount{affil:Monard2}{
     Kleinkaroo Observatory, Center for Backyard Astronomy Kleinkaroo,
     Sint Helena 1B, PO Box 281, Calitzdorp 6660, South Africa}



\maketitle

\begin{abstract}
We identified that ASASSN-14cc is a very active dwarf nova
spending approximately 60\% of the time in outburst.
Our long-term photometry revealed that the object shows
long outbursts recurring with a period of 21--33~d
and very brief short outbursts lasting less than 1~d.
The maximum decline rate exceeds 2.8 mag d$^{-1}$.
The duration of long outbursts is 9--18~d, comprising
50--60\% of the recurrence time of long outbursts.
We detected 0.01560--0.01562~d (22.5 min) modulations
during long outbursts, which we identified to be superhumps.
These features indicate that ASASSN-14cc has 
outburst parameters very similar to the extreme
dwarf nova RZ LMi but with a much shorter superhump
period.  All the observations can be naturally understood
considering that this object is a helium analog
(AM CVn-type object) of RZ LMi.  The highest outburst
activity among AM CVn-type objects can be understood
as the high-mass transfer rate expected for the
orbital period giving a condition close to the stability
limit of the accretion disk.  In contrast to RZ LMi,
this object shows little evidence for premature
quenching of the superoutburst, which has been proposed
to explain the unusual outburst parameters in RZ LMi.
\end{abstract}

\section{Introduction}

   Cataclysmic variable (CVs) are close binaries consisting
of a white dwarf accreting and a red-dwarf secondary 
transferring matter via the Roche-lobe overflow.
Although many of the CVs have a low-mass, hydrogen-rich secondary
star, some CVs have a helium white dwarf as the secondary
and such systems are called AM CVn-type objects
[for recent reviews of AM CVn-type objects,
see e.g. \citet{nel05amcvnreview}; \citet{sol10amcvnreview}].
Some AM CVn-type objects show dwarf nova-type outbursts
and they are considered to be a result of thermal instability
in the helium accretion disk \citep{tsu97amcvn}.
These outbursts are considered to be generally analogous
to those of hydrogen-rich systems [for a review of CVs in general,
see e.g. \citet{war95book}; \citet{hel01book} and a review
for dwarf nova outburst, see \citet{osa96review}].
It has been established that AM CVn-type objects also
show a sequence of superoutbursts and normal outbursts,
as in SU UMa-type dwarf novae in hydrogen-rich systems
[e.g. \citet{kat00crboo}; see \citet{osa89suuma} for
thermal-tidal instability model of SU UMa-type dwarf novae].
The outburst phenomenon in AM CVn-type objects is now
under intensive research (e.g. \cite{lev13amcvnPTF};
\cite{lev15amcvn}; \cite{Pdot4}) and many new objects have been
identified.  A comparison between hydrogen-rich CVs
and AM CVn-type systems may provide a clue in understanding
the difference in physics of outbursts in 
hydrogen and helium disks.

   Here we report on a discovery of a very unusual object,
ASASSN-14cc, which we propose to be a helium analog of
the very unusual hydrogen-rich CV, RZ LMi.

\section{ASASSN-14cc}

   ASASSN-14cc is a transient discovered by 
the ASAS-SN \citep{ASASSN} team on 2014 June 8 at $V$=15.8.\footnote{
  $<$http://www.astronomy.ohio-state.edu/$\sim$assassin/transients.html$>$.
}
The coordinates are
\timeform{21h 39m 48.24s}, \timeform{-59D 59' 32.4''}
(USNO-B1.0 counterpart).
The object was readily recognized as a frequently outbursting
object in the Catalina Real-time Transient Survey (CRTS)
public database.\footnote{
  $<$http://nesssi.cacr.caltech.edu/DataRelease/$>$.
}
We realized that the object was recorded in outburst
(brighter than 17.7 mag) in 96 observations out of
156 observations between JD 2453592 (2005 August 9)
and 2456408 (2013 April 25).  The unusually high outburst
duty cycle of 62\% urged us to monitor this object.

\section{Observation and Analysis}\label{sec:obs}

   The observations were performed by two observers
between 2014 June 11 and 2014 December 14:
F.-J. Hambsch (HMB, 40cm telescope in San Pedro de Atacama, Chile;
\cite{ham12ROADaavso})
and B. Monard (MLF, 35cm telescope Center for Backyard Astronomy
Kleinkaroo Observatory).  Both observers used snapshot modes
mainly in quiescence and time-series photometry during outbursts.
The exposure times were 90s by HMB and 30s by MLF.
The sampling rate of time-series photometry by HMB was typically
one frame per one to five minutes.  The dead time between
exposures by MLF is typically 1~s.
Both observers used unfiltered photometry.
After standard de-biasing and flat-fielding, the brightness
of the object was measured by aperture photometry.
By using nearby comparison and check stars
(HMB used GSC 9103.1409 for comparison and GSC 8819.1006 for check;
MLF used GSC 8819.981 for comparison and GSC 2.2 S3111120464
for check), the magnitudes
were converted to the $V$ system, which approximates well
the response of unfiltered CCDs for outbursting dwarf novae
($B-V \sim$0).  The total numbers of observations were
3235 on 162 nights (HMB) and 1927 on 11 nights (MLF).

   We have converted the times of observations into
barycentric Julian days (BJDs) before the analysis.
We used phase dispersion minimization (PDM; \cite{PDM})
for period analysis and 1$\sigma$ errors for the PDM analysis
were estimated by the methods of \citet{fer89error} and \citet{Pdot2}.
For de-trending the global variations, we used
locally-weighted polynomial regression (LOWESS: \cite{LOWESS}).

\section{Results}\label{sec:result}

\subsection{Outburst}\label{sec:out}

\begin{figure*}
  \begin{center}
    \FigureFile(170mm,85mm){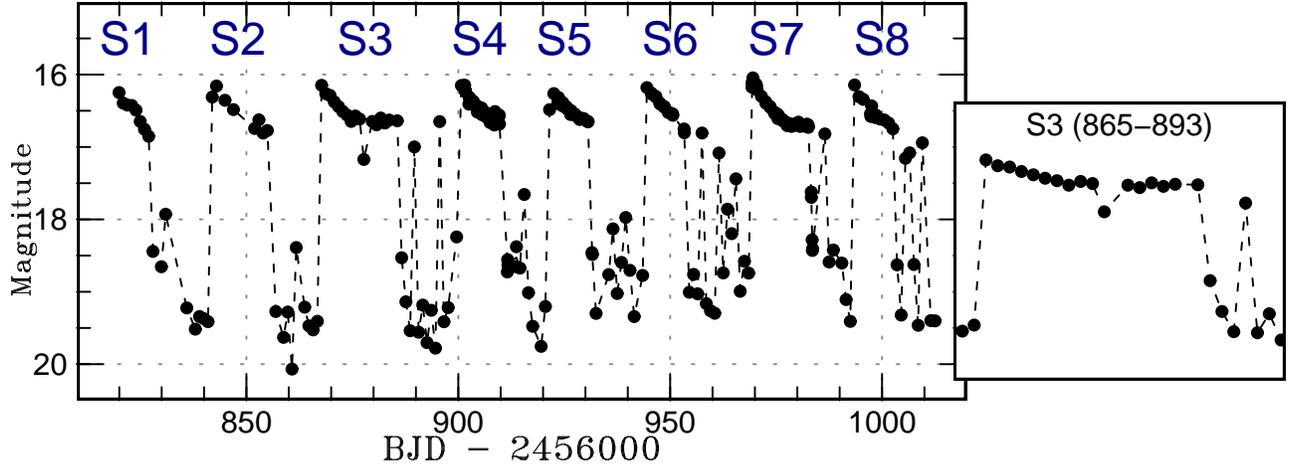}
  \end{center}
  \caption{Long-term light curve of ASASSN-14cc.  The data
  were binned to 0.1 d.  Eight long outbursts (superoutbursts,
  marked as S1--S8) and intervening low state with short
  outbursts are clearly seen.  The light behavior is very
  similar to RZ LMi.  The dashed lines are added to aid
  recognizing the variation.  The right panel is an enlargement
  of S3.}
  \label{fig:sn14cclc}
\end{figure*}

   The overall light curve of ASASSN-14cc is shown in figure
\ref{fig:sn14cclc}.  Eight long outbursts reaching
16.0--16.3 mag were recorded and these long outbursts
were intervened by low states usually reaching 19.0--20.0 mag,
but the low states became brighter (around 19.0 mag)
in the later part of our observation.
During the low states, extremely short (less than 1~d)
outbursts were recorded.  These features resemble
those of the most unusual SU UMa-type dwarf nova RZ LMi
(\cite{rob95eruma}; \cite{nog95rzlmi};
\cite{ole08rzlmi}) but the duration of the short outbursts
is extremely short compared to RZ LMi.  During two long
outbursts, there were short dips at BJD 2456877.7
and 2457003.5--2457004.5, which were not observed in RZ LMi.
The fastest fading rate
reached at least 2.8 mag d$^{-1}$ (such as between
BJD 2456895.6 and BJD 2456896.6).  Since snapshot
observations were taken only once a night, the true
fading rates of short outbursts should have exceeded
this value.
This feature is also extreme compared to RZ LMi.

   The starting dates (BJD$-$2456000) of the long outbursts were
841.9 (S2), 867.8 (S3), 900.7 (S4), 921.5 (S5), 944.5 (S6),
969.5 (S7) and 993.5 (S8) allowing $\sim$0.5~d errors due to 
gaps of observations.  The intervals of long outbursts 
were variable between 21~d and 33~d.
The durations of these outbursts
were 13--14~d (S2), 18~d (S3, including a dip), 
9--10~d (S4), 11~d (S5), 9~d (S6), 14~d (S7) and 
13~d (S8, including a dip).\footnote{
  We have identified a fading on 1003.5--1004.5 as a dip
because the object stayed bright again on two nights
(1005.5 and 1006.5) after this fading.
}
These long outbursts compose nearly 50--60\% of
the intervals, confirming the high value of duty cycle
suggested by the CRTS data.

\subsection{Superhump}\label{sec:superhump}

\begin{figure}
  \begin{center}
    \FigureFile(88mm,70mm){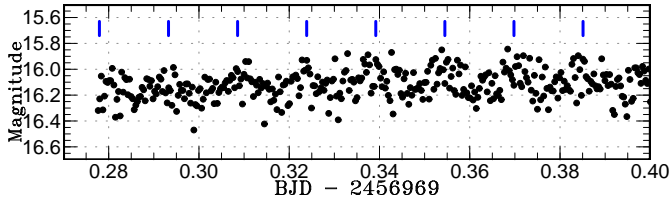}
  \end{center}
  \caption{Example of time-resolved photometry of ASASSN-14cc.
  The observations were obtained by MLF on the first night
  of the outburst S7.  The epochs of superhump maxima are
  labeled by vertical ticks.}
  \label{fig:sn14ccsh}
\end{figure}

\begin{figure}
  \begin{center}
    \FigureFile(88mm,110mm){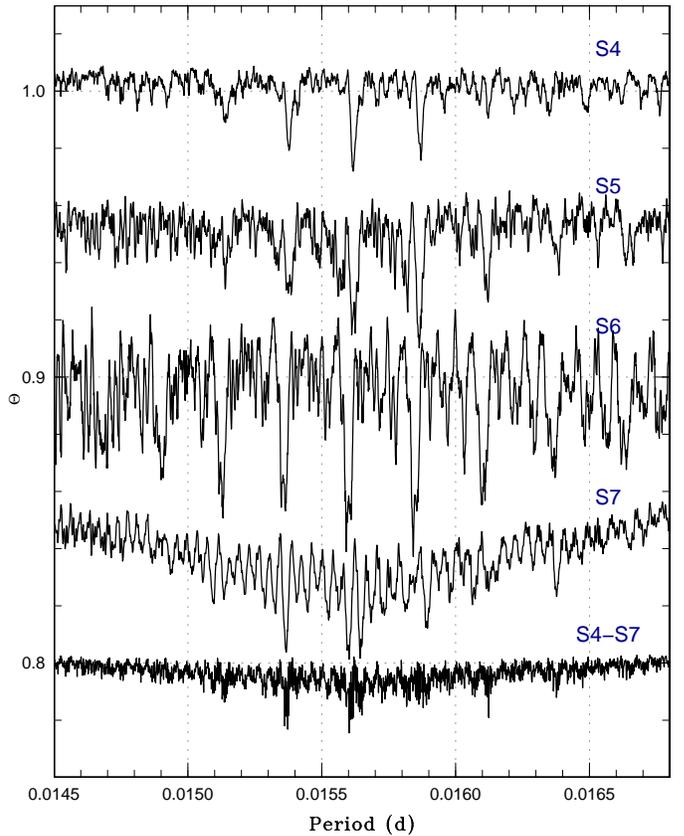}
  \end{center}
  \caption{PDM analysis of long outbursts in ASASSN-14cc.
  The period of 0.01560--0.01562 d (see text for details)
  is detected in all long outbursts.
  The $\Theta$ values are the real values for S4 and
  values for the others were shifted by 0.05
  between outbursts.
  An analysis of the combined data is shown at the bottom. 
  }
  \label{fig:sn14ccpdmcomp}
\end{figure}

   We performed time-series photometry during four
long outbursts S4--S8.  On the first night of S7,
the existence of hump signals were already visible
in the light curve (figure \ref{fig:sn14ccsh}).
A PDM analysis of this segment of the data yielded
a period of 0.01564(5)~d with a mean amplitude
of 0.13 mag.  Although the hump signals were
weaker in the later part of the long outbursts,
and the time-resolution of the observations by HMB
was not high enough to depict the hump signals
directly, the signal with the same period was
consistently detected in four long outbursts
(figure \ref{fig:sn14ccpdmcomp}).

\begin{figure}
  \begin{center}
    \FigureFile(88mm,110mm){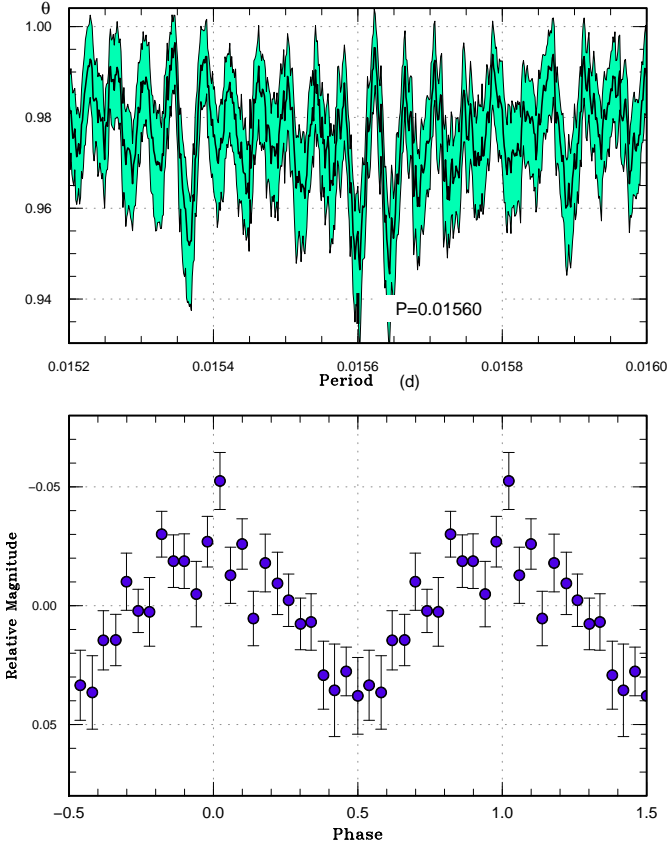}
  \end{center}
  \caption{Superhumps in ASASSN-14cc.
  (Upper): PDM analysis of the outburst S7 (interval
  before BJD 2456976.  We analyzed 100 samples which randomly
  contain 70\% of observations, and performed PDM analysis for
  these samples.  The bootstrap result is shown as a form of 90\%
  confidence intervals in the resultant $\theta$ statistics.
  (Lower): Phase-averaged profile.}
  \label{fig:sn14ccshpdm}
\end{figure}

   The best periods for long outbursts are summarized
in table \ref{tab:shper}.  In this table, the entire
duration of the outburst was used for analysis
for S4 and S5.  The analysis was limited to the early
part with better signal-to-noise ratio for S6
(BJD before 2456951) and S7 (BJD before 2456976;
see figure \ref{fig:sn14ccshpdm} for the analysis
in full resolution).
The subtle difference in the period and amplitude
may reflect the difference of the interval used
for analysis.  The outburst S8 was affected by
strong moonlight and no meaningful period was
obtained.  This outburst was not included in the following
combined analysis.

   An analysis of the combined data for S4--S7
(bottom curve of figure \ref{fig:sn14ccpdmcomp})
yielded a period of 0.015602(1)~d.  There is a possibility
that these modulations are coherent between
different long outbursts.  This issue will be
discussed in subsection \ref{sec:coherence}.

   We identified these modulations as superhumps
in analogy with RZ LMi (\cite{rob95eruma}; \cite{nog95rzlmi};
\cite{ole08rzlmi}), although the identification
of superhumps in RZ LMi was still based on
phenomenology (strong variations during long, bright
outbursts) and not by the secure detection of
the orbital period (see e.g. \cite{Pdot4} for the
current situation).  These variations were found to be
strongest during the initial part of the long outburst
(as seen in S7), which also agrees with the trend
of superhumps in general.  Following the identification
of superhumps, we call these long outbursts
superoutbursts.

\begin{table}
\caption{Superhump periods.}\label{tab:shper}
\begin{center}
\begin{tabular}{ccc}
\hline
Outburst & Period (d) & Amplitude (mag) \\
\hline
S4 & 0.015621(6) & 0.05 \\
S5 & 0.015620(7) & 0.05 \\
S6 & 0.015602(8) & 0.06 \\
S7 & 0.015600(2) & 0.09 \\
\hline
\end{tabular}
\end{center}
\end{table}

\section{Discussion}\label{sec:discuss}

\subsection{ASASSN-14cc as a Likely Helium Analog of RZ LMi}\label{sec:rzlmianalog}

   As we have seen in section \ref{sec:superhump},
ASASSN-14cc has all the properties of the extreme
SU UMa-type dwarf nova RZ LMi.
The major difference is in the superhump period
(the superhump period in RZ LMi is around 0.0594~d,
cf. \cite{Pdot4}).
The short superhump period in ASASSN-14cc
cannot be explained by a hydrogen-rich ordinary CV.
The only currently feasible interpretation is
an AM CVn-type object with a helium white dwarf
as the secondary.  Although there remains a possibility
that this object may contain some hydrogen, as in
EI Psc-type objects with a stripped core evolved secondary
(e.g. \cite{uem02j2329letter}; \cite{tho02j2329};
Ohshima et al. in preparation), the shortest known
superhump period in this class of object 
(0.0382~d in CRTS J102842.9$-$081927: \cite{Pdot}; \cite{Pdot4};
\cite{Pdot5}; \cite{wou12SDSSCRTSCVs}) is much longer
than in ASASSN-14cc, and we consider this possibility
less likely.

   Although this identification must be confirmed
by spectroscopy, we continue the discussion on
this supposed identification.  Since we do not know
the exact orbital period, we use the superhump period
(usually different from the orbital period by $\sim$1\%
in AM CVn-type objects) as the approximate orbital period
in the following discussion.

\subsection{ASASSN-14cc on Activity Sequence of AM CVn-Type Objects}\label{sec:activity}

   It is known that outburst activity in AM CVn-type
objects is strongly dependent on orbital periods
(cf. \cite{nel05amcvnreview}; \cite{lev15amcvn}).
Objects with orbital periods shorter than $\sim$20 min
do not show dwarf nova-type outbursts and they are
equivalent to novalike objects in hydrogen-rich systems.
Objects with orbital periods 20--40 min generally show
dwarf nova-type outbursts and objects with longer orbital
periods generally do not show outbursts.  Very recently,
two objects having orbital periods 47--48 min have been
found to show outbursts (SDSS J090221.35$+$381941.9:
\cite{kat14j0902}; CSS J045019.7$-$093113: \cite{wou13j0450atel4726}).
This activity sequence can be naturally understood
by the disk instability model and assuming that
the mass-transfer in AM CVn-type objects is powered
by gravitational wave radiation \citep{tsu97amcvn}.

   It is very interesting that the period (22.5 min)
of ASASSN-14cc is located on the lower border of
dwarf nova-type AM CVn-type objects.  It implies that
the object is located on the borderline between
a thermally stable disk (novalike object) and
a thermally unstable disk (dwarf nova-type object)
and the observed behavior is in very good agreement
with this expectation.  Among known AM CVn-type objects,
PTF1 J191905.19$+$481506.2 (hereafter PTF1 J1919)
shows most active
dwarf nova-type outbursts [its supercycle, which is the cycle length
between superoutbursts, is 36.8(4)~d] and it has an orbital period
of 22.4559(3) min \citep{lev14j1919}.  The case is more
extreme in ASASSN-14cc showing a supercycle as short as
21~d with a much higher value of the duty cycle of the superoutburst
compared to PTF1 J1919.

\subsection{Implication on Disk-Instability Model}\label{sec:DImodel}

   In hydrogen-rich systems, it is well-known that a higher
mass-transfer rate produces a shorter supercycle.
There is, however, a shortest limit since the addition
of the transferred matter produces a longer superoutburst
(cf. \cite{osa95eruma}).  The expected limit is $\sim$43~d
for hydrogen-rich systems and ER UMa is exactly such
an object (\cite{kat95eruma}; \cite{osa95eruma}).
RZ LMi, however, has a shorter supercycle of 19~d,
which cannot be reproduced by a simple increase of
the mass-transfer rate.  \citet{osa95rzlmi} proposed
that a larger disk radius at the end of the superoutburst
can explain such a short supercycle.  This radius
corresponds to the strength of the tidal torque
and \citet{osa95rzlmi} suggested that a weaker tidal
torque is responsible for the phenomenon.
\citet{hel01eruma} also explained the phenomenon
in the same context.

   It is not known what is the expected shortest limit
of a supercycle in AM CVn-type dwarf novae
without a modification of the disk radius at
the end of the superoutburst.
Our observation indicates a higher value of
the duty cycle (0.5--0.6) compared to the model of
RZ LMi in \citet{osa95rzlmi}.
The outburst light curve resembles that of a model
light curve with a mass-transfer rate higher than
ER UMa without a modification in the disk radius
(upper panel of figure 3 in \cite{osa95eruma}).
These findings suggest that premature quenching of
superoutbursts do not occur in this system.
Since the mass ratios (and hence the strength
of the tidal torque) are expected to be smaller
in AM CVn-type objects than in hydrogen-rich systems,
the observation may suggest that a small mass ratio
is not a sufficient explanation for the RZ LMi
phenomenon.

   Although currently there is no parameter study
for helium dwarf novae covering the highest mass-transfer
rates, such a study will shed light on the
interpretation of ER UMa/RZ LMi-type phenomenon
in hydrogen-rich and helium CVs.

\subsection{Coherence of Superhump Signals between
Different Superoutbursts}\label{sec:coherence}

   It has been claimed in RZ LMi that superhump signals
were coherent between different superoutbursts
\citep{ole08rzlmi}.  This coherence of the signals
has raised important questions: (1) whether these modulations
are indeed superhumps, (2) whether superhumps are continuously
excited, which may be supportive evidence for
the interpretation by \citet{hel01eruma}.

\begin{figure}
  \begin{center}
    \FigureFile(75mm,95mm){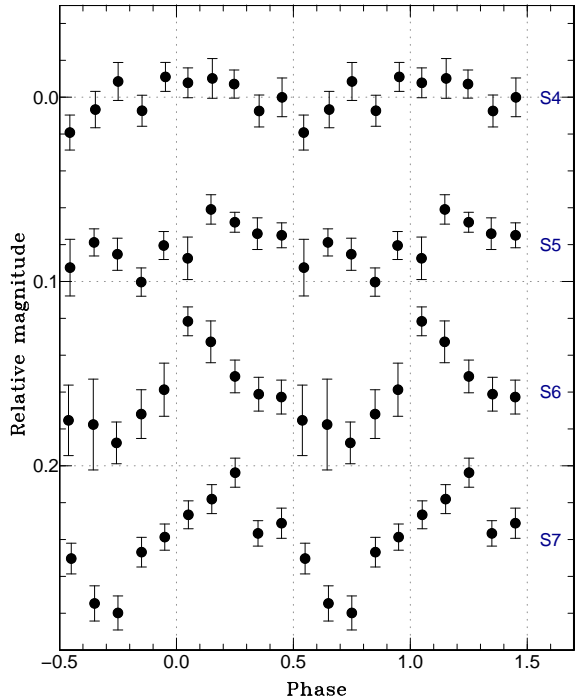}
  \end{center}
  \caption{Phase-averaged profile of superhumps in ASASSN-14cc.
  An ephemeris of BJD(max)$=2456900+0.015602 E$ was used.
  The superhump maxima were not coherent in phase between
  different superoutbursts.
  }
  \label{fig:sn14ccphasecomp}
\end{figure}

   ASASSN-14cc may be a suitable object for discussing
this issue, since we have observations for four consecutive
superoutbursts.  There was only a single viable period
common to four superoutbursts (subsection \ref{sec:superhump})
and figure \ref{fig:sn14ccphasecomp} represents phase-averaged
light curves of superhumps against this period.
The result clearly indicates that superhumps are not
coherent between different superoutbursts, and
these signals cannot be the orbital signal. 
These observations suggest that superhumps are newly
excited at the onset of each superoutburst, which is
in agreement with the expectation for the thermal-tidal
disk instability model
(\cite{osa89suuma}; \cite{osa95eruma}).

\section*{Acknowledgements}

This work was supported by the Grant-in-Aid
``Initiative for High-Dimensional Data-Driven Science through Deepening
of Sparse Modeling'' from the Ministry of Education, Culture, Sports, 
Science and Technology (MEXT) of Japan.
We are grateful to the ASAS-SN team for making their
detections of CV outbursts available to the public.
We are grateful to the Catalina Real-time Transient Survey
team for making their database available to the public.
We thank K. Isogai for helping in the compilation
of the observation.

\end{document}